\documentclass[showpacs,10pt,twocolumn,prb]{revtex4-1}

\usepackage{amsmath}
\usepackage{amssymb}
\usepackage{graphicx}
\usepackage{amssymb}
\usepackage{graphics}
\usepackage{epsfig}
\usepackage{CJK}
\usepackage{color}

\begin{document}

\begin{CJK*}{GBK}{Song}
\title{Anisotropic magnetocaloric effect and critical behavior in CrCl$_3$}
\author{Yu Liu and C. Petrovic}
\affiliation{Condensed Matter Physics and Materials Science Department, Brookhaven National Laboratory, Upton, New York 11973, USA}
\date{\today}

\begin{abstract}
We report anisotropic magnetocaloric effect and magnetic critical behavior in van der Waals crystal CrCl$_3$. The maximum magnetic entropy change $-\Delta S_M^{max} \sim 14.6$ J kg$^{-1}$ K$^{-1}$ and the relative cooling power $RCP \sim 340.3$ J kg$^{-1}$ near $T_c$ with a magnetic field change of 5 T are much larger when compared to CrI$_{3}$ or CrBr$_{3}$. The rescaled $\Delta S_M(T,H)$ curves collapse onto a universal curve, confirming the second order ferromagnetic transition. Further critical behavior analysis around $T_c$ presents a set of critical exponents $\beta$ = 0.28(1) with $T_c$ = 19.4(2) K, $\gamma$ = 0.89(1) with $T_c$ = 18.95(8) K, and $\delta$ = 4.6(1) at $T_c$ = 19 K, which are close to those of theoretical tricritical mean field model.
\end{abstract}
\maketitle
\end{CJK*}

\section{INTRODUCTION}

Chromium trihalides CrX$_3$ (X = Cl, Br, and I) have triggered a renewed interest since the recent discovery of intrinsic two-dimensional (2D) magnetism in monolayer CrI$_3$.\cite{Huang} These materials provide platform for engineering of novel spintronic devices and studies of 2D spin order. The microscopic mechanism for properties of interest stems from the layered antiferromagnetic (AFM) ground state and the low critical magnetic fields required for a ferromagnetic (FM) phase transition (6 kOe and 11 kOe for bilayer CrI$_3$ and CrCl$_3$, respectively).\cite{Huang,McGuire0,Seyler,Gong,Huang0,Huang1,Jiang,Cai}

Bulk CrI$_3$ and CrBr$_3$ are FM with the Curie temperature ($T_c$) of 61 K and 37 K,\cite{McGuire,Dillon,Tsu} respectively, whereas CrCl$_3$ is AFM with the N\'{e}el temperature ($T_N$) of 16.8 K.\cite{Cable} Bulk CrX$_3$ crystalize in the layered BiI$_3$-type structure, space group R$\bar{3}$. The edge-shared CrX$_6$ octahedra form a 2D honeycomb layers of Cr ions. These sandwiched X-Cr-X slabs are stacked along the $c$ axis and are held by weak van der Waals (vdW) interactions. The different radii of X alter the in-plane nearest-neighbor Cr-Cr distance and the vdW gap between X-Cr-X slabs. From I to Br to Cl, the X-Cr-X bonding is less covalent, weakening superexchange interactions and lowering ordering temperature.\cite{McGuire0} For CrCl$_3$, neutron scattering and NMR experiments show a three-dimensional (3D) AFM magnetic structure at low temperature, consisting of alternating FM sheets of spins aligned within the Cr planes.\cite{Cable} The transition temperature of 17 K was characterized by heat capacity measurement,\cite{Hansen,Starr} and recently updated with two heat capacity peaks at $T_N$ = 14 K and $T_C$ = 17 K.\cite{McGuire0} Faraday rotation, magnetization, and neutron diffraction measurements show that the ordered state of CrCl$_3$ has a weak magnetic anisotropy, and fields of only a few kOe are required to fully polarize the magnetization in- or out-of the Cr plane.\cite{Kuhlow,Bizette,Bykovetz} The magnetocaloric effect (MCE) of vdW magnets can give additional insight into the magnetic properties, and it can also be used to assess magnetic refrigeration potential. Bulk CrI$_3$ exhibits anisotropic magnetic entropy change ($-\Delta S_M^{max}$) with values of 4.24 and 2.68 J kg$^{-1}$ K$^{-1}$ at 50 kOe for out-of-plane and in-plane fields, respectively.\cite{YuLIU} The value of $-\Delta S_M^{max}$ is about 7.2 J kg$^{-1}$ K$^{-1}$ at 50 kOe for CrBr$_3$.\cite{Xiao} Typical 3D magnetic critical behavior is present in CrI$_3$ crystals.\cite{LIUYU,Gao} However, the magnetocaloric properties and critical behavior of CrCl$_3$ are still unknown.

In the present work we investigate the unusual two-step magnetic ordering process of bulk CrCl$_3$ single crystals by detailed measurements of dc and ac magnetization. The AFM ground state below $T_N$ = 14.4 K is observed, and an intermediate FM before transition into PM state on heating is also confirmed. The values of $-\Delta S_M^{max} \sim 14.6$ J kg$^{-1}$ K$^{-1}$ and the relative cooling power ($RCP) \sim 340.3$ J kg$^{-1}$ near the PM-FM phase transition with field change of 5 T, indicating that CrCl$_3$ would be a promising candidate material for cryomagnetic refrigeration. The scaling analysis of $\Delta S_M(T,H)$ reveals that the PM-FM phase transition is of second-order in nature. A set of critical exponents is further estimated, $\beta$ = 0.26(1), $\gamma$ = 0.86(1), and $\delta$ = 4.6(1), indicating that the PM-FM transition at $T_c$ $\sim$ 19 K of bulk CrCl$_3$ is situated close to a 3D to 2D critical point.

\section{EXPERIMENTAL DETAILS}

CrCl$_3$ single crystals were grown by recrystalizing commercial CrCl$_3$ (Alpha Aesar, 99.9$\%$) polycrystal using the chemical vapor transport (CVT) method. The starting material was sealed in a quartz tube in vacuum and then placed inside a two-zone horizontal tube furnace with source and growth temperatures up to 650 $^\circ$C and 550 $^\circ$C, respectively, for 7 days. Large, thin, violet-colored, transparent plate-like single crystals with lateral dimensions up to several millimeters can be obtained. The x-ray diffraction (XRD) data were taken with Cu K$_{\alpha}$ ($\lambda=0.15418$ nm) radiation of Rigaku Miniflex powder diffractometer. The magnetic properties were characterized by the quantum design magnetic property measurement system (MPMS-XL5). The dc magnetization was measured at various magnetic fields from 5 Oe to 50 kOe. The isothermals were measured up to 50 kOe in $\Delta T$ = 1 K intervals.

\section{RESULTS AND DISCUSSIONS}

\begin{figure}
\centerline{\includegraphics[scale=1]{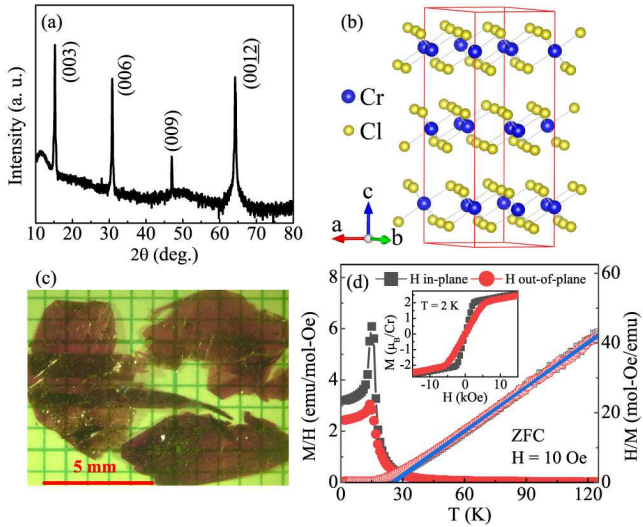}}
\caption{(Color online). (a) X-ray diffraction (XRD) pattern in log scale of CrCl$_3$ single crystal. BiI$_3$-type structure (b) at low temperature and representative single crystals (c). (d) Temperature dependence of zero field cooling (ZFC) normalized magnetization (left axis) $M/H$ and (right axis) $H/M$ of CrCl$_3$ at H = 10 Oe applied along in-plane and out-of-plane directions. Inset shows the field dependence of magnetization $M(H)$ at 2 K.}
\label{1}
\end{figure}

The XRD pattern can be well indexed by the indices of $(00l)$ plane, indicating that the crystal surface is parallel to the $ab$ plane [Figs. 1(a-c)]. The in-plane and out-of-plane directions are parallel and perpendicular to the $ab$ plane, respectively. It should be noted that CrX$_3$ share similar structural transitions from the low-temperature rhombohedral to high-temperature monoclinic symmetry with CrX$_{3}$: 210 K for CrI$_3$, 420 K for CrBr$_3$, and 230 K for CrCl$_3$, respectively.\cite{Handy} The interlayer spacing $d$ = 5.8 {\AA} is calculated from the peak positions using Bragg$^\prime$s law $n\lambda = 2dsin\theta$, consistent with the previously reported value.\cite{Abramchuk} Here we focus on the low-temperature magnetic properties of CrCl$_3$. Figure 1(d) shows the temperature dependence of normalized magnetization $M/H$ at $H$ = 10 Oe applied parallel and perpendicular to the $ab$ plane as well as the inverse values with a temperature step of 1 K. Since there is no significant difference of the zero-field-cooling (ZFC) and field-cooling (FC) data only the ZFC curves are presented. The $M/H$ curves show a peak near 15 K for both field directions. The downturn below 15 K points to the reported AFM ground state, while the rapid increase just above it hints towards a possible FM intermediate state. The high temperature data can be fitted by the Curie-Weiss law, giving the effective magnetic moment of 4.32(2)/4.37(2) $\mu_B$/Cr, somewhat larger than the expected value 3.87 $\mu_B$ for spin-only Cr$^{3+}$ ion, and the Weiss temperature of 26.1(3)/24.5(2) K for in-plane/out-of-plane field. The positive Weiss temperature indicates that the FM interactions dominate the magnetic behavior in the paramagnetic (PM) state. The isothermal magnetization of CrCl$_3$ at 2 K for both field directions shows negligible hysteresis, as shown in inset of Fig. 1(d). The in-plane data change to a weak field-dependence at a smaller field compared to the out-of-plane data indicating in-plane easy axis and a weak anisotropy. The linear increase at low fields shows characteristic of the behavior expected for polarizing an antiferromagnet with weak anisotropy.

\begin{figure}
\centerline{\includegraphics[scale=1]{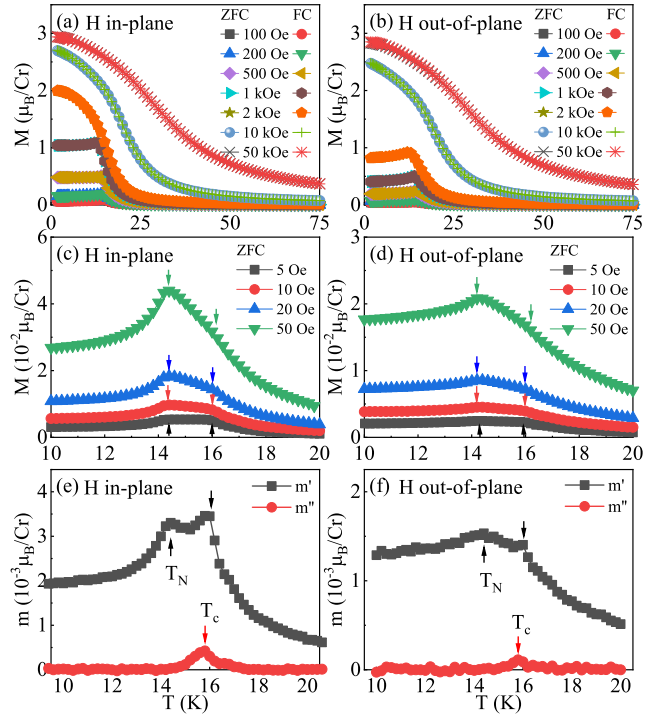}}
\caption{(Color online). Temperature dependence of zero field cooling (ZFC) and field cooling (FC) dc magnetization $M(T,H)$ of CrCl$_3$ measured at the indicated (a,c) in-plane and (b,d) out-of-plane magnetic fields. Temperature dependence of ac susceptibility real part $m^\prime(T)$ and imaginary part $m^{\prime\prime}(T)$ measured with oscillated ac field of 3.8 Oe and frequency of 499 Hz applied (e) in-plane and (f) out-of-plane.}
\label{2}
\end{figure}

The temperature dependence of the magnetic moment per Cr measured at various fields near the magnetic transition is shown in Figs. 2(a) and 2(b). At 2 K and 50 kOe a moment of 3.0 $\mu_B$/Cr is obtained, as expected for Cr$^{3+}$ with S = 3/2. The curves are very similar to those reported in Ref. 2. Kuhlow reported similar behavior in Faraday-rotation measurements,\cite{Kuhlow} which was interpreted as 2D FM order within the layers developing first, with interlayer long-range AFM order setting in at lower temperature. In order to characterize this two-step magnetic ordering process in CrCl$_3$, we further present the $M(T)$ data at low fields below 100 Oe with a temperature step of 0.2 K [Figs. 2(c) and 2(d)]. Two distinct peaks in 14.4 K and 16.0 K are clearly observed in low-field dc magnetization, as well as in the real part $m'$ of ac magnetization [Figs. 2(e) and 2(f)] . The imaginary part $m''$ feature a peak anomaly at 16.0 K but not at 14.4 K, further confirming that AFM ground state stems from a FM-like immediate state.

\begin{figure}
\centerline{\includegraphics[scale=1]{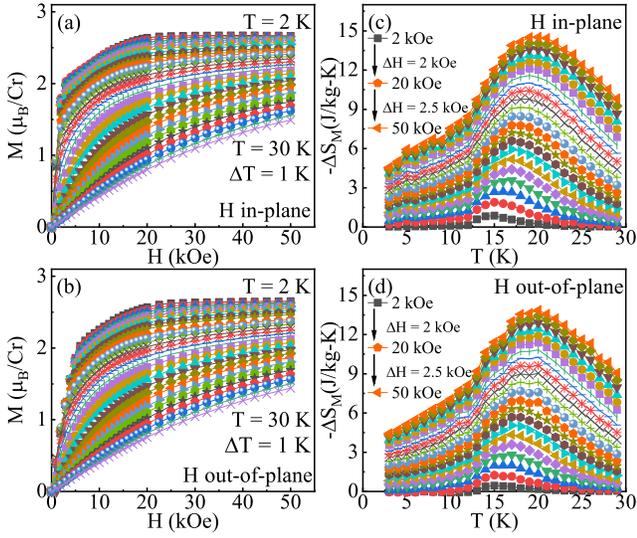}}
\caption{(Color online). Representative magnetization isothermals at various temperatures around $T_c$ for (a) in-plane and (b) out-of-plane magnetic fields. Temperature dependence of the derived magnetic entropy change $-\Delta S_M$ at various (c) in-plane and (d) out-of-plane magnetic fields.}
\label{3}
\end{figure}

Figures 3(a) and 3(b) exhibit the isothermal magnetization with field up to 50 kOe applied along in-plane and out-of-plane directions, respectively, from 2 K to 30 K with a temperature step of 1 K. At high temperature, the curves are almost linear, suggesting a PM behavior. With decreasing temperature, the curves bend with negative curvatures, indicating an FM interaction. At low temperature, there is a rapid linear increase at low field and the magnetic moment is fully polarized in the high-field region. Based on the classical thermodynamical and the Maxwell's thermodynamic relation, the magnetic entropy change $\Delta S_M(T,H)$ is given by:\cite{Pecharsky,Amaral}
\begin{equation}
\Delta S_M = \int_0^H [\frac{\partial S(T,H)}{\partial H}]_TdH = \int_0^H [\frac{\partial M(T,H)}{\partial T}]_HdH,
\end{equation}
where $[\partial S(T,H)/\partial H]_T$ = $[\partial M(T,H)\partial T]_H$ is based on the Maxwell's relation. For magnetization measured at small temperature and field intervals,
\begin{equation}
\Delta S_M = \frac{\int_0^HM(T_{i+1},H)dH-\int_0^HM(T_i,H)dH}{T_{i+1}-T_i}.
\end{equation}
The calculated $-\Delta S_M(T,H)$ are presented in Figs. 3(a) and 3(b). All the curves exhibit a peak feature, and the peak broadens asymmetrically on both sides with increasing magnetic field. For both field directions, the peak position gradually shifts from 15 K for 2 kOe to 19 K for 50 kOe. The $-\Delta S_M$ reaches to a maximum value $\sim$ 14.6 J kg$^{-1}$ K$^{-1}$ for in-plane field and 13.8 J kg$^{-1}$ K$^{-1}$ for out-of-plane field, respectively.

\begin{figure}
\centerline{\includegraphics[scale=1]{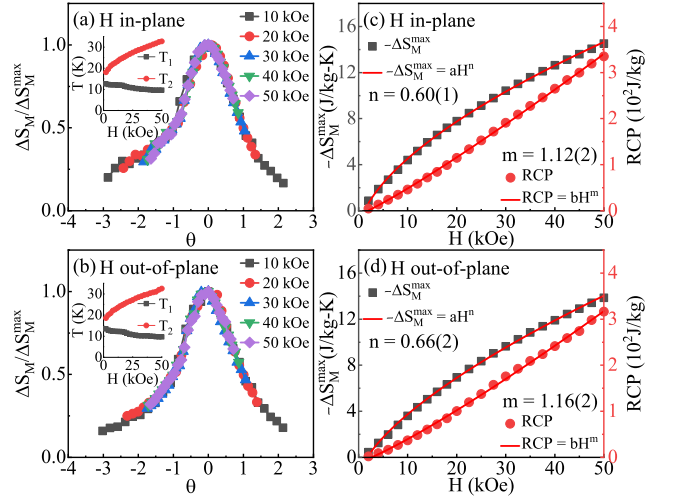}}
\caption{(Color online). Normalized $\Delta S_M/\Delta S_M^{max}$ as a function of the rescaled temperature $\theta$  for (a) in-plane and (b) out-of-plane fields. Insets show the evolution of the reference temperatures $T_1$ and $T_2$. Field dependence of the maximum magnetic entropy change $-\Delta S_M^{max}$ and the relative cooling power RCP with power law fitting in red solid lines for (c) in-plane and (d) out-of-plane fields.}
\label{4}
\end{figure}

There is a generalized magnetic entropy scaling analysis proposed for the second-order phase transition magnetocaloric materials.\cite{Franco} In this approach the normalized magnetic entropy $\Delta S_M/\Delta S_M^{max}$, estimated for each constant magnetic field, is scaled to the reduced temperature $\theta$ as defined in the following equations:
\begin{equation}
\theta_- = (T_{peak}-T)/(T_{r1}-T_{peak}), T<T_{peak},
\end{equation}
\begin{equation}
\theta_+ = (T-T_{peak})/(T_{r2}-T_{peak}), T>T_{peak},
\end{equation}
where $T_{r1}$ and $T_{r2}$ are the lower and upper temperatures at full-width half maximum of $\Delta S_M/\Delta S_M^{max}$. In this method, $T_c$ fails to be a good parameter whereas $T_{peak}$ serves the purpose because of its field dependence. The normalized $\Delta S_M/\Delta S_M^{max}$ roughly collapses on to a universal curve around $T_{peak}$ at indicated fields [Figs. 4(a) and 4(b)], indicating the feature of second-order PM-FM transition in CrCl$_3$. Another parameter to characterize the potential magnetocaloric effect of materials is the relative cooling power (RCP):\cite{Gschneidner}
\begin{equation}
RCP = -\Delta S_M^{max} \times \delta T_{FWHM},
\end{equation}
where the FWHM means the full width at half maximum of $-\Delta S_M$ curve. The RCP reaches a maximum value of 340.3 J kg$^{-1}$ for in-plane field and 317.3 J kg$^{-1}$ for out-of-plane field, respectively. In addition, the field dependence of $-\Delta S_M^{max}$ and RCP can be well fitted by using the power-law relations $-\Delta S_M^{max} = aH^n$ and $RCP = bH^m$ [Figs. 4(c) and 4(d)].\cite{VFranco} The $-\Delta S_M^{max}$ of CrCl$_3$ is smaller than that of well-known magnetic refrigerating materials with first-order transition,\cite{Gschneidner1} but is larger than some with second-order transition.\cite{MH1,MH2,MH3} For instance, the $-\Delta S_M^{max}$ of R$_6$Co$_{1.67}$Si$_3$ (R = Pr, Gd, and Tb) are 6.9, 5.2 and 7.0 J kg$^{-1}$ K$^{-1}$ at 50 kOe, and for CdCr$_2$S$_4$ is 7.0 J kg$^{-1}$ K$^{-1}$ at the same field.\cite{Shen,Tishin} It is worth noting that the values of $-\Delta S_M^{max}$ and RCP of CrCl$_3$ are also larger than those of its cousin CrBr$_3$ (7.2 J kg$^{-1}$ K$^{-1}$ and 191.5 J kg$^{-1}$) and CrI$_3$ (4.24 J kg$^{-1}$ K$^{-1}$ and 122.6 J kg$^{-1}$).\cite{YuLIU,Xiao} Thus, bulk CrCl$_3$ could be a promising candidate for cryogenic magnetic refrigerating materials.

\begin{figure}
\centerline{\includegraphics[scale=1]{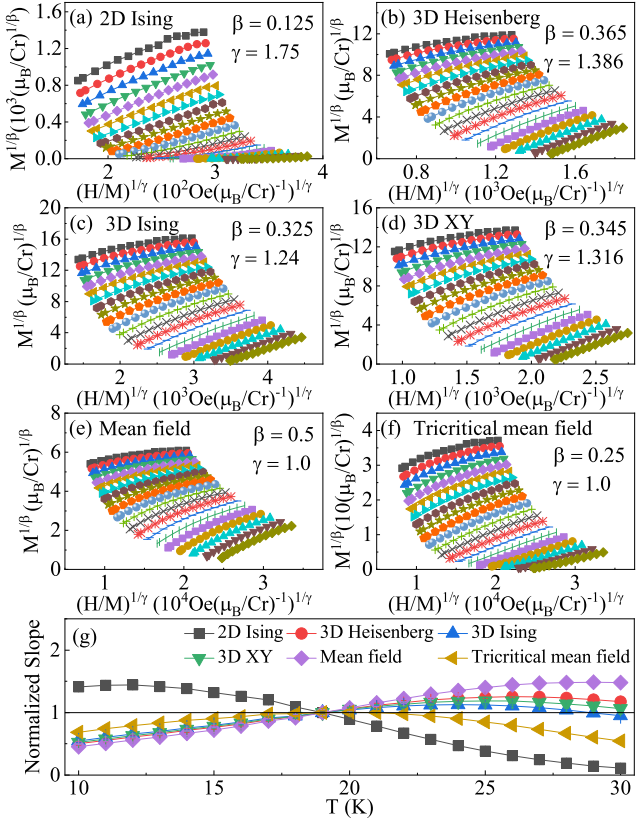}}
\caption{(Color online). The modified Arrott plots $M^{1/\beta}$ vs $(H/M)^{1/\gamma}$ for in-plane fields with parameters of (a) 2D Ising, (b) 3D Heisenberg, (c) 3D Ising, (d) 3D XY, (e) Mean field, and (f) Tricritical mean field models. (g) Temperature dependence of the normalized slopes $NS = S(T)/S(T_c)$ for different theoretical models.}
\label{5}
\end{figure}

\begin{figure}
\centerline{\includegraphics[scale=1]{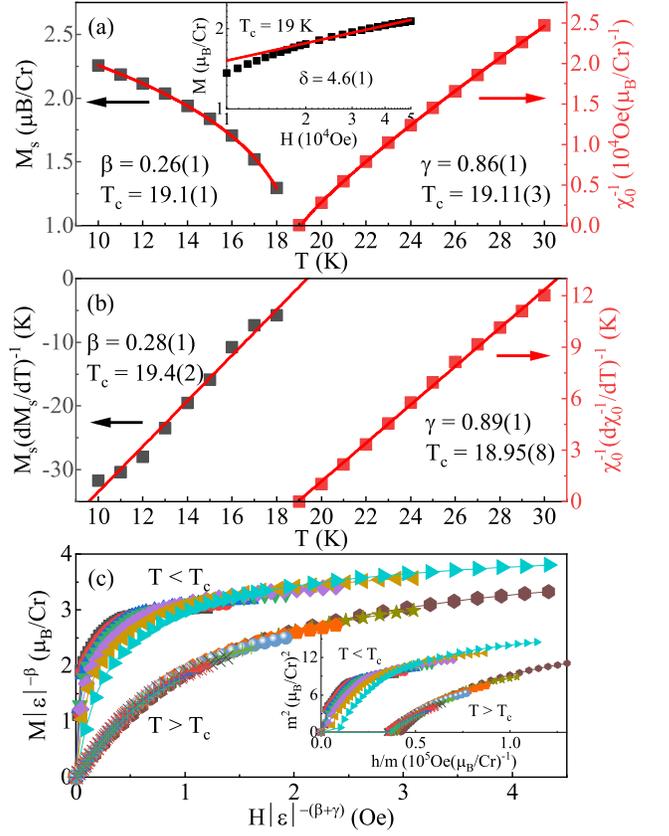}}
\caption{(Color online). (a) Temperature dependence of the spontaneous magnetization $M_s$ (left) and the inverse initial susceptibility $\chi_0^{-1}$ (right) with solid fitting curves for CrCl$_3$. Inset shows the log$_{10}M$ vs log$_{10}H$ at $T_c$ = 19 K with linear fitting curve. (b) Kouvel-Fisher plots of $M_s(dM_s/dT)^{-1}$ (left) and $\chi_0^{-1}(d\chi_0^{-1}/dT)^{-1}$ (right) with solid fitting curves for CrCl$_3$. (c) Scaling plots of renormalized $m = M|\varepsilon|^{-\beta}$ vs $h = H|\varepsilon|^{-(\beta+\gamma)}$ above and below $T_c$ for CrCl$_3$. Inset shows the Arrott plot of $m^2$ vs $h/m$ for CrCl$_3$.}
\label{6}
\end{figure}

For a second-order PM-FM phase transition, the spontaneous magnetization ($M_s$) below $T_c$, the initial magnetic susceptibility ($\chi_0^{-1}$) above $T_c$, and the field-dependent magnetization (M) at $T_c$ are characterized by a set of critical exponents $\beta$, $\gamma$, and $\delta$, respectively.\cite{Stanley} The mathematical definitions of the exponents from magnetization measurement are given below:
\begin{equation}
M_s (T) = M_0(-\varepsilon)^\beta, \varepsilon < 0, T < T_c,
\end{equation}
\begin{equation}
\chi_0^{-1} (T) = (h_0/m_0)\varepsilon^\gamma, \varepsilon > 0, T > T_c,
\end{equation}
\begin{equation}
M = DH^{1/\delta}, T = T_c,
\end{equation}
where $\varepsilon = (T-T_c)/T_c$; $M_0$, $h_0/m_0$ and $D$ are the critical amplitudes.\cite{Fisher} For the original Arrott plot, $\beta$ = 0.5 and $\gamma$ = 1.0.\cite{Arrott1} In a more general case with different critical exponents, the Arrott-Noaks equation of state provides a modified Arrott plot:\cite{Arrott2}
\begin{equation}
(H/M)^{1/\gamma} = a\varepsilon+bM^{1/\beta},
\end{equation}
where $\varepsilon = (T-T_c)/T_c$ and $a$ and $b$ are fitting constants. Figures 5(a)-5(f) present the modified Arrott plots for easy in-plane fields using theoretical critical exponents from 2D Ising ($\beta = 0.125, \gamma = 1.75$), 3D Heisenberg ($\beta = 0.365, \gamma = 1.386$), 3D Ising ($\beta = 0.325, \gamma = 1.24$), 3D XY ($\beta = 0.345, \gamma = 1.316$), mean-field ($\beta = 0.5, \gamma = 1.0$) and tricritical mean-field ($\beta = 0.25, \gamma = 1.0$) models.\cite{Kaul,Khuang,LeGuillou} There should be a set of parallel lines in the high fields with the same slope $S(T) = dM^{1/\beta}/d(H/M)^{1/\gamma}$. Comparing the normalized slope [$NS = S(T)/S(T_c)$] with the ideal value of 1 enables us to determine the most suitable model, as shown in Fig. 5(g). It is clearly seen that the $NS$ of 2D Ising model shows the largest deviation from 1. The $NS$ of tricritical mean field model is close to $NS = 1$ mostly below $T_c$, while that of 3D Ising model is the best above $T_c$.

To generate the actual critical exponents of bulk CrCl$_3$, the linearly extrapolated $M_s$ and $\chi_0^{-1}$ are plotted against temperature in Fig. 6(a).\cite{Pramanik} According to Eqs. (6) and (7), the solid fitting lines give that $\beta = 0.26(1)$, with $T_c = 19.1(1)$ K, and $\gamma = 0.86(1)$, with $T_c = 19.11(3)$ K. According to Eq. (8), the $M(H)$ at $T_c$ should be a straight line in log-log scale with the slope of $1/\delta$. Such fitting yields $\delta = 4.6(1)$ [inset in Fig. 6(a)], which agrees with the calculated $\delta = 4.3(1)$ from the obtained $\beta$ and $\gamma$ based on the Widom relation $\delta = 1+\gamma/\beta$.\cite{Widom} The more accurate values of critical exponents could be obtained using the Kouvel-Fisher technique, where $M_s(T)/(dM_s(T)/dT)^{-1}$ and $\chi_0^{-1}(T)/(d\chi_0^{-1}(T)/dT)^{-1}$ plotted against temperature should be straight lines with slopes $1/\beta$ and $1/\gamma$, respectively.\cite{Kouvel} The linear fits to the plots [Fig. 6(b)] yield the values of critical exponents and $T_c$ are $\beta = 0.28(1)$, with $T_c = 19.4(2)$ K, and $\gamma = 0.89(1)$, with $T_c = 18.95(8)$ K. The value of $\beta$ for a 2D magnet should be within a window $0.1 \leq \beta \leq 0.25$,\cite{Taroni} suggesting a possible 3D magnetic behavior of CrCl$_3$. The obtained values of critical exponents are close to those of theoretical tricritical mean field model ($\beta$ = 0.25 and $\gamma$ = 1.0), indicating that the second-order PM-FM transition is situated close to a 3D to 2D critical point.

Scaling analysis can be used to estimate the reliability of the obtained critical exponents. According to scaling hypothesis, the magnetic equation of state in the critical region obeys a scaling relation can be expressed as:
\begin{equation}
M(H,\varepsilon) = \varepsilon^\beta f_\pm(H/\varepsilon^{\beta+\gamma}),
\end{equation}
where $f_+$ for $T > T_c$ and $f_-$ for $T < T_c$, respectively, are the regular functions. In terms of the variable $m\equiv\varepsilon^{-\beta}M(H,\varepsilon)$ and $h\equiv\varepsilon^{-(\beta+\gamma)}H$, scaled or renormalized magnetization and scaled or renormalized field, respectively, Eq.(10) reduces to a simple form:
\begin{equation}
m = f_\pm(h).
\end{equation}
It implies that for a true scaling relation with proper selection of $\beta$, $\gamma$, and $\delta$, the scaled $m$ versus $h$ data will fall onto two different universal curves; $f_+(h)$ for temperature above $T_c$ and $f_-(h)$ for temperature below $T_c$. Using the values of  $\beta$ and $\gamma$ obtained from the Kouvel-Fisher plot, we have constructed the scaled $m$ vs scaled $h$ plot in Fig. 6(c). It is clear from the plots that all the experimental data collapse onto two different branches: one above $T_c$ and another below $T_c$. The deviation at low fields below $T_c$ probably arise from the spin dynamics behavior of CrCl$_3$. The scaling analysis can be also verified from plots of $m^2$ vs $h/m$ [inset in Fig. 6(c)], confirming proper treatment of the critical regime.

\section{CONCLUSIONS}

In summary, we have studied in details the magnetism and magnetocaloric effect of bulk CrCl$_3$ single crystal. The two-step magnetic transition at $T_N$ = 14.4 K and $T_c$ = 16 K was clearly characterized by low-field dc and ac magnetization measurements. Further neutron scattering measurement is of interest to shed more light on its microscopic mechanism. The magnetic entropy change $-\Delta S_M^{max} \sim 14.6$ J kg$^{-1}$ K$^{-1}$ and the relative cooling power $RCP \sim 340.3$ J kg$^{-1}$ with in-plane field change of 50 kOe indicates that CrCl$_3$ would be a promising candidate material for cryomagnetic refrigeration. The second-order in nature of the PM-FM transition near $T_c$ has been verified by the scaling analysis of $-\Delta S_M$. A set of critical exponents $\beta$, $\gamma$, and $\delta$ estimated from various techniques match reasonably well and follow the scaling equation, indicating that the PM-FM transition of bulk CrCl$_3$ is situated close to a 3D to 2D critical point. Considering its magnetism can be maintained upon exfoliating bulk crystal down to a monolayer,\cite{Besbes,Xue,MacNeill,Pocs,Kazim,Lu} further study on the size-dependent properties is of interest.

\section*{Acknowledgements}
This work was supported by the US DOE-BES, Division of Materials Science and Engineering, under Contract No. DE-SC0012704 (BNL).

\end{document}